\documentclass[a4paper, 9pt]{extarticle}
\usepackage[raggedright]{titlesec}
\usepackage{amsmath,amssymb,amsfonts}
\usepackage{graphicx,color}
\usepackage{times}
\usepackage{tikz}
\usepackage{multicol}
\usepackage{ccicons}
\usepackage{fancyhdr}
\usepackage[utf8]{inputenc}
 \usepackage[english]{babel}

\fancyhfoffset{0cm}
\titleformat{\section}[hang]
  {\usefont{T1}{qhv}{b}{n}}
  {}
  {0em}
  {\hspace{-0.4pt}\Large \thesection\hspace{0.6em}}

\titleformat{name=\section,numberless}[hang]
  {\usefont{T1}{qhv}{b}{n}}
  {}
  {0em}
  {\hspace{-0.4pt}\Large}

\usepackage[font={footnotesize,bf}]{caption}
\usepackage{floatrow} 
\usepackage{hyperref} 
\usepackage{etoolbox} 
\patchcmd{\thebibliography}{\section*{\refname}}{}{}{}

\hypersetup{
		pdftitle={Predicting Natural Hazards with Neuronal Networks},    
		pdfauthor={Matti und Dani},     
		pdfsubject={Predicting Natural Hazards with Neuronal Networks},   
		colorlinks=true,
		linkcolor=[rgb]{0, 0, 0},          
		citecolor=[rgb]{0.41,0.53,0.71},        
		filecolor=[rgb]{0.74 0.42 0.34},      
		urlcolor=[rgb]{0.41,0.53,0.71},         
}

\definecolor{col1}{HTML}{6988B5}
\definecolor{col2}{HTML}{516168}
\definecolor{col3}{HTML}{945C5B}
\definecolor{col4}{HTML}{BE6C56}
\definecolor{col5}{HTML}{FA9C56}
\definecolor{col6}{HTML}{FEC967}
\definecolor{col7}{HTML}{FFE77D}
\definecolor{col8}{HTML}{CFCE3C}

\usetikzlibrary{decorations.pathreplacing}
\usetikzlibrary{calc,3d,shadings}
\usetikzlibrary{arrows.meta}
\usetikzlibrary{fadings}
\usetikzlibrary{calc}

\renewcommand{\r}[1]{\mathrm{#1}}
\renewcommand{\b}[1]{\mathbf{#1}}
\newcommand{\bs}[1]{\boldsymbol{#1}}

\date{}

\title{\large\textbf{Predicting Natural Hazards with Neuronal Networks}}
\author{\normalsize Matthias Rauter$^{1,2}$ \qquad Daniel Winkler$^{3}$\\[1ex]
\small \textit{$^1$ University of Innsbruck, Unit of Geotechnical and Tunnel Engineering, \href{mailto:matthias.rauter@uibk.ac.at}{matthias.rauter@uibk.ac.at}}\\
\small \textit{$^2$ Austrian Research Centre for Forests, Department of Natural Hazards}\\
\small \textit{$^3$ University of Innsbruck, Unit of Environmental Engineering, \href{mailto:daniel.winkler@uibk.ac.at}{daniel.winkler@uibk.ac.at}}\\
}

\begin{document}

\maketitle

\section*{Abstract}

Gravitational mass flows, such as avalanches, debris flows and rockfalls are common events in alpine regions with high impact on transport routes. Within the last few decades, hazard zone maps have been developed to systematically approach this threat. These maps mark vulnerable zones in habitable areas to allow effective planning of hazard mitigation measures and development of settlements. Hazard zone maps have shown to be an effective tool to reduce fatalities during extreme events. They are created in a complex process, based on experience, empirical models, physical simulations and historical data. The generation of such maps is therefore expensive and limited to crucially important regions, e.g.\ permanently inhabited areas. 

In this work we interpret the task of hazard zone mapping as a classification problem. Every point in a specific area has to be classified according to its vulnerability. On a regional scale this leads to a segmentation problem, where the total area has to be divided in the respective hazard zones. The recent developments in artificial intelligence, namely convolutional neuronal networks, have led to major improvement in a very similar task, image classification and semantic segmentation, i.e.\ computer vision. 
We use a convolutional neuronal network to identify terrain formations with the potential for catastrophic snow avalanches and label points in their reach as vulnerable. Repeating this procedure for all points allows us to generate an artificial hazard zone map. We demonstrate that the approach is feasible and promising based on the hazard zone map of the Tirolean Oberland. However, more training data and further improvement of the method is required before such techniques can be applied reliably.

\section{Introduction}

\paragraph{Hazard Zone Maps}
Natural Hazards, particularly gravitational mass flows are constant threats to settlement and infrastructure in alpine regions. Beside fatalities and destroyed buildings, such events can lead to blocked or destroyed transport routes. To mitigate the impact of natural hazards, transport routes and settlement areas are protected by various artificial barriers, such as dams, avalanche galleries, and snow fences. 

\begin{figure}[htb]
\includegraphics[width=\textwidth]{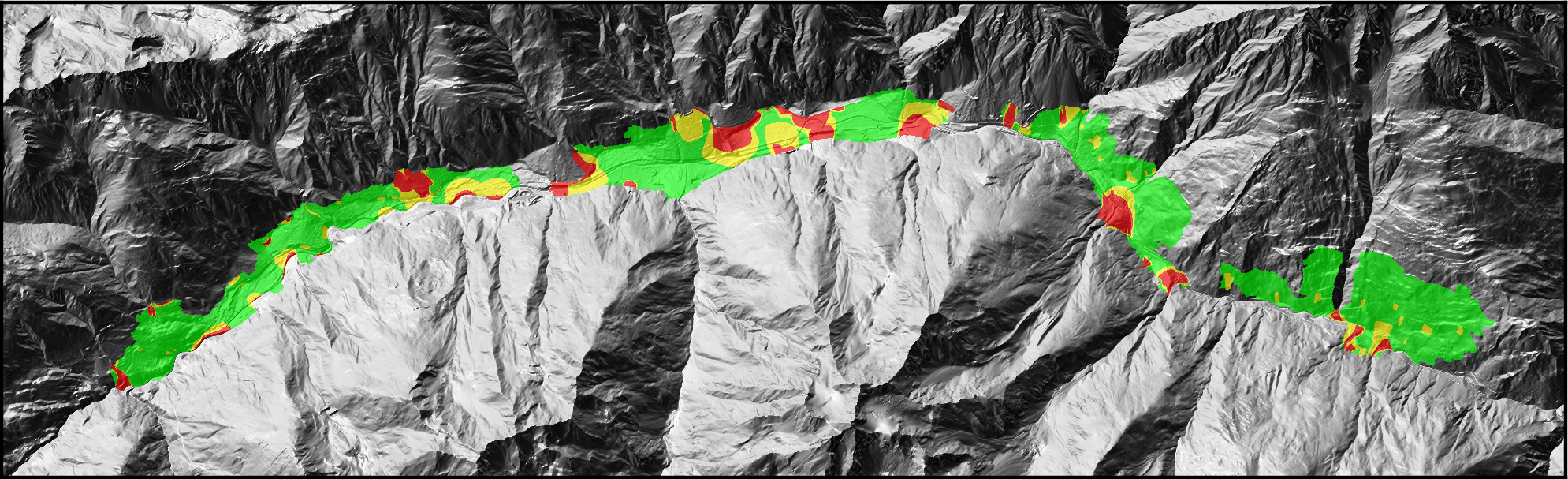}
\caption{The hazard zone map regarding snow avalanches of the Stanzer Valley.}
\label{fig:gzp}
\end{figure}

The essential basis for planning protection measures and settlement development are hazard zone maps. They have been introduced in 1975 in Austria \cite{BMLFUW2011} and in 1997 in South Tirol \cite{BZ2012}. In Austria, these maps mark every point in habitable areas as vulnerable (yellow) or highly vulnerable (red). These categories and colour codes are slightly different in other countries (e.g. Switzerland and South Tirol) but the principle is the same. In here we use an additional colour (green) to mark areas which have been identified as safe, see Figure~\ref{fig:gzp}. Uncoloured areas have not been subject of a detailed analysis and show the terrain (hill shade). Hazard zone maps are developed in a complex and expensive process. Therefore there are many areas where such a detailed analysis is missing.

Although understanding and mitigation strategies have improved significantly within the last decades, we still struggle to handle natural disasters regularly. We are continuously reminded of this fact by disasters like the snow avalanche in Farindola (Abruzzo) or the debris flow in Puster Valley (South Tirol), to name two events from the year 2017. This highlights the demand for improved models and procedures to (1) improve predictions and reduce uncertainties and (2) extend the investigated area.

\paragraph{Artificial Intelligence and Neuronal Networks}

Neuronal networks are one of the most common approaches for artificial intelligence and machine learning. This technique is applied when a task is too complex to develop an algorithm for its solution. The principal idea is to develop an application which learns to solve the problem on its own. Often, this is simpler than solving the problem directly. This approach has shown outstanding performance for notoriously difficult tasks, such as image and speech recognition \cite{schmidhuber2015deep}. 


With our work we aim to explore the possibilities of neuronal networks for the management of natural hazards. 
Neuronal networks can be used to develop temporal and spatial models which are learned autonomously from historical data (one might say they gather experience). In fact, this approach makes it possible to process historical data and human expertise to make suggestions for future decisions. In other words, neuronal networks learn from past catastrophes and from experienced engineers. From this point of view, the neuronal network is just a statistical-empirical model, similar to the well established $\alpha$-$\beta$-model \cite{lied1980empirical}. Both, the $\alpha$-$\beta$-model and our neuronal network, are processing topographic features. However, the neuronal network has to extract these features from terrain data (elevation maps) on its own without any human help. An advantage of this approach is that we do not have to choose specific terrain features beforehand, the network chooses them based on statistical considerations. This allows the neuronal network to work mostly autonomously. Obviously, there should be human supervision but the goal is to reduce the human effort to the essential tasks. We hope that neuronal networks help to improve natural hazard mitigation and therefore safety standards in alpine regions. Also, neuronal networks may reduce costs to make detailed studies and hazard zone maps available in more regions (e.g.\ Abruzzo) and not permanently inhabited areas (e.g.\ transport routes).

There have been some attempts to process geographical data like hazard zones with neuronal networks. Some groups (e.g.\ Lee et al. \cite{lee2004determination}) performed landslide susceptibility assessments with neuronal networks with encouraging results. However, these approaches are not directly comparable to our network, because important features, such as slope inclination and curvature are extracted manually from elevation maps. A more comparable study on geographical scale with outstanding results is shown by Isola et al.\ \cite{isola2016image}, automatically transforming satellite images to street maps with neuronal networks. Our idea is similar, transforming terrain maps to hazard zone maps.

In this work we focus on snow avalanches. Snow avalanches are extremely complex and notoriously difficult to predict. There is a high demand for new and improved models. Moreover, we focus on learning and generating hazard zone maps. Hazard zone maps do not change with time in contrast to momentary reports like the avalanche bulletin. Therefore, we focus on the recognition of spatial features which are decisive for catastrophic snow avalanches. To train neuronal networks, one needs examples of input data and the corresponding output data. Usually humans have to create databases with such examples to allow the neuronal network to learn. Terrain data in combination with the official hazard zone maps act as human generated training data for our purpose. Using these maps, the neuronal network can learn from experience, history and all other inputs used for the hazard zone maps.

We expect that this approach is also applicable to other natural hazards, such as floods and debris flows. Although it may be required to adapt the chosen network architecture and input data to the specific problem, this work can act as blueprint for these hazards.

This paper is organised as follows: In Section 2 we outline state of the art neuronal networks. In Section 3 we present our implementation, specialised in predicting snow avalanches. In Section 4 we describe the training phase and show results on training and validation samples. Finally, we summarise the work in Section 5 and highlight our plans for the future. 

\section{Neuronal Networks in a Nutshell}

\begin{figure}[htb]
\includegraphics[width=0.8\textwidth]{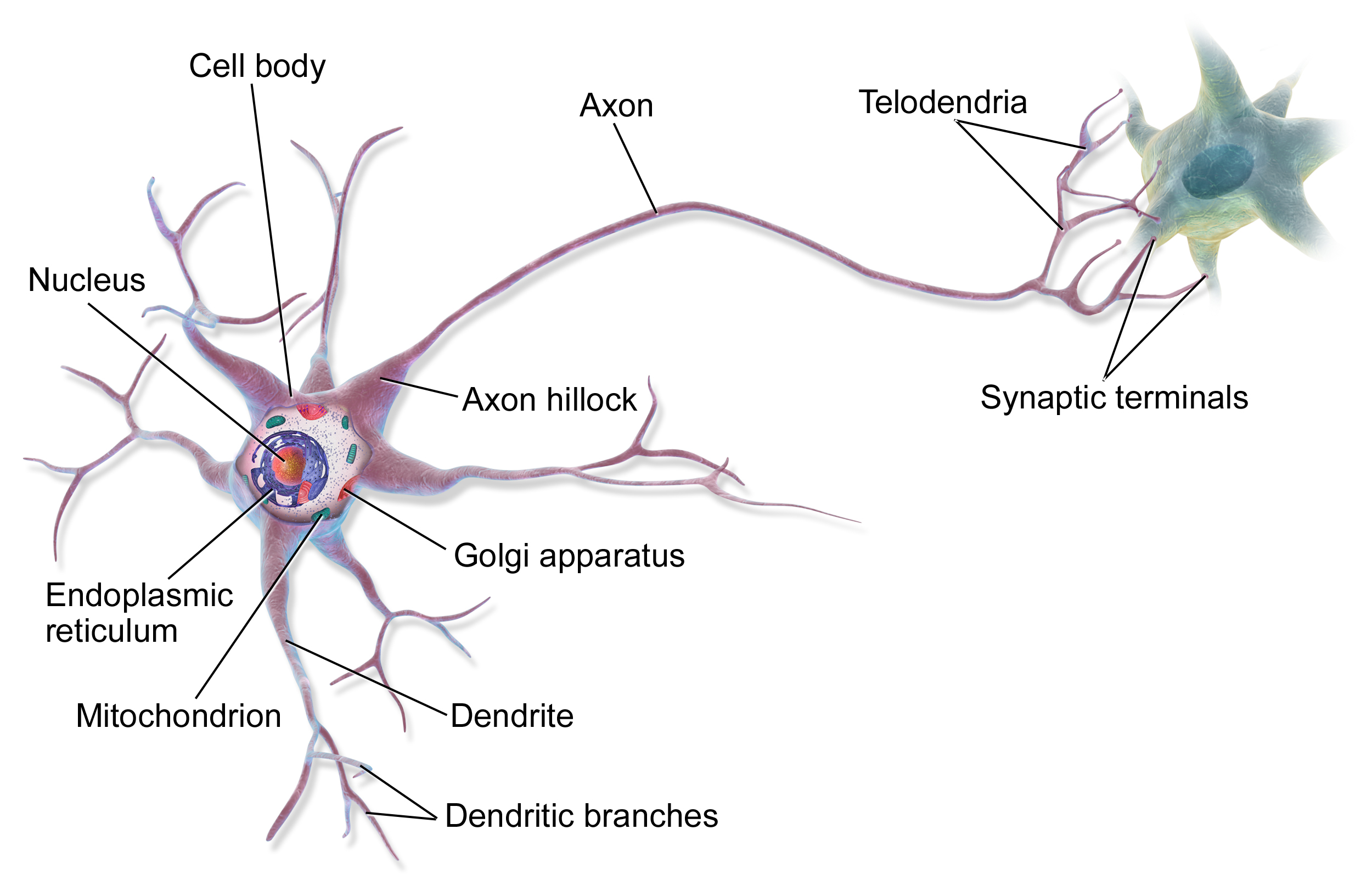}
\caption{A single neuron. Picture: Blausen Medical Communications, Inc. \ccby}
\label{fig:neuron}
\end{figure}

\begin{figure}[htb]
\begin{tikzpicture}[yscale=-1]
\node[circle, draw, inner sep=1.5pt] (sum) at (3.5,1.6) {$\sum$};
\node[circle, draw, inner sep=3.5pt] (bias) at (3.5,0) {$b$};
\node[circle, draw, inner sep=2.5pt] (fun) at (5.1,1.6) {$f$};

\node[inner sep=3pt] (in1) at (0,0) {$x_1$};
\node[inner sep=3pt] (in2) at (0,0.8) {$x_2$};
\node[inner sep=3pt] (in3) at (0,1.6) {$x_3$};
\node[inner sep=3pt] (in5) at (0,3.2) {$x_n$};

\node[circle, draw, inner sep=1.5pt, xshift=1.8cm, yshift=0cm] (w1) at (in1) {$w_1$};
\node[circle, draw, inner sep=1.5pt, xshift=1.8cm, yshift=0cm] (w2) at (in2) {$w_2$};
\node[circle, draw, inner sep=1.5pt, xshift=1.8cm, yshift=0cm] (w3) at (in3) {$w_3$};
\node[circle, draw, inner sep=1.5pt, xshift=1.8cm, yshift=0cm] (w5) at (in5) {$w_n$};

\draw[-{Latex[length=2mm,width=1.2mm]}] (in1) -- (w1);
\draw[-{Latex[length=2mm,width=1.2mm]}] (in2) -- (w2);
\draw[-{Latex[length=2mm,width=1.2mm]}] (in3) -- (w3);
\draw[-{Latex[length=2mm,width=1.2mm]}] (in5) -- (w5);

\draw[-{Latex[length=2mm,width=1.2mm]}] (w1) -- (sum);
\draw[-{Latex[length=2mm,width=1.2mm]}] (w2) -- (sum);
\draw[-{Latex[length=2mm,width=1.2mm]}] (w3) -- (sum);
\draw[-{Latex[length=2mm,width=1.2mm]}] (w5) -- (sum);

\node[inner sep=0.2pt, fill=black] at (0, 2.2) {};
\node[inner sep=0.2pt, fill=black] at (0, 2.4) {};
\node[inner sep=0.2pt, fill=black] at (0, 2.6) {};

\draw[-{Latex[length=2mm,width=1.2mm]}] (bias) -- (sum);
\draw[-{Latex[length=2mm,width=1.2mm]}] (sum) -- (fun);
\draw[-{Latex[length=2mm,width=1.2mm]}] (fun) -- +(1,0) node[right] {$y$};

\draw [decorate,decoration={brace,amplitude=10pt},xshift=-4pt,yshift=0pt]
(5.5, 3.5) -- (1.5, 3.5) node [circle, draw, inner sep=2.0pt,black,midway,yshift=-25pt] {\tikz{\draw[] (0,0) ..controls (0.25,0) and (0.15,0.3) .. (0.4,0.3);}};

\end{tikzpicture}
\caption{McCulloc-Pits-cell for the simulation of a single neuron. A complete neuron is later on marked by a node with a wave in its middle.}
\label{fig:mcculloc-pits}
\end{figure}
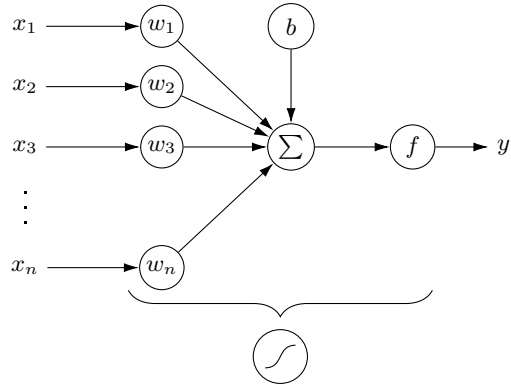

The origin of neuronal networks can be traced back to the early work of McCulloc and Pits \cite{mcculloch1943logical}, who first introduced a mathematical model for a biological neuron. A neuron is stimulated through its dendrites, the signal is processed and a new stimulus is passed through the axon to the dendrites of other neurons (see Figure~\ref{fig:neuron}). The McCulloc-Pits-cell is a mathematical description of this process. The processing of the signal is implemented in different ways. The standard implementation consists of a linear weighting, adding a bias and a nonlinear activation function. This mechanism is illustrated in Figure~\ref{fig:mcculloc-pits}. It can also be expressed by the simple function
\begin{align}
y = f\left(\sum\limits_i\,w_{i}\,x_i + b\right),
\end{align}
where $y$ is the output of the neuron and $x_i$ the $i$th input. $f$ is the activation function, usually a rectified linear unit or sigmoid function. This function introduces some non-linearity to the network, increasing the complexity of its behaviour. The weights $w_i$ determine on which stimulus the cell reacts, the bias $b$ acts as threshold, which has to be overcome before the cell emits a stimulus on its own. The weights $w_{i}$ and the bias $b$ are mutable to allow the cell to change its behaviour in the learning phase.

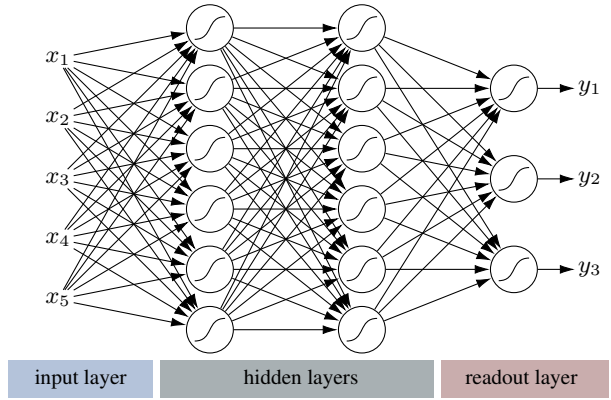
\begin{figure}[htb]
\begin{tikzpicture}[yscale=-1]

\node[inner sep=1pt] (in1) at (0,0) {$x_1$};
\node[inner sep=1pt] (in2) at (0,0.8) {$x_2$};
\node[inner sep=1pt] (in3) at (0,1.6) {$x_3$};
\node[inner sep=1pt] (in4) at (0,2.4) {$x_4$};
\node[inner sep=1pt] (in5) at (0,3.2) {$x_5$};

\begin{scope}[xshift=2.0cm]
\node[circle, draw, inner sep=1pt] (w1) at (0, -0.4) {\tikz{\draw[] (0,0) ..controls (0.25,0) and (0.15,0.3) .. (0.4,0.3);}};
\node[circle, draw, inner sep=1pt] (w2) at (0,  0.4) {\tikz{\draw[] (0,0) ..controls (0.25,0) and (0.15,0.3) .. (0.4,0.3);}};
\node[circle, draw, inner sep=1pt] (w3) at (0,  1.2) {\tikz{\draw[] (0,0) ..controls (0.25,0) and (0.15,0.3) .. (0.4,0.3);}};
\node[circle, draw, inner sep=1pt] (w4) at (0,  2.0) {\tikz{\draw[] (0,0) ..controls (0.25,0) and (0.15,0.3) .. (0.4,0.3);}};
\node[circle, draw, inner sep=1pt] (w5) at (0,  2.8) {\tikz{\draw[] (0,0) ..controls (0.25,0) and (0.15,0.3) .. (0.4,0.3);}};
\node[circle, draw, inner sep=1pt] (w6) at (0,  3.6) {\tikz{\draw[] (0,0) ..controls (0.25,0) and (0.15,0.3) .. (0.4,0.3);}};
\end{scope}

\begin{scope}[xshift=4.0cm]
\node[circle, draw, inner sep=1pt] (v1) at (0, -0.4) {\tikz{\draw[] (0,0) ..controls (0.25,0) and (0.15,0.3) .. (0.4,0.3);}};
\node[circle, draw, inner sep=1pt] (v2) at (0,  0.4) {\tikz{\draw[] (0,0) ..controls (0.25,0) and (0.15,0.3) .. (0.4,0.3);}};
\node[circle, draw, inner sep=1pt] (v3) at (0,  1.2) {\tikz{\draw[] (0,0) ..controls (0.25,0) and (0.15,0.3) .. (0.4,0.3);}};
\node[circle, draw, inner sep=1pt] (v4) at (0,  2.0) {\tikz{\draw[] (0,0) ..controls (0.25,0) and (0.15,0.3) .. (0.4,0.3);}};
\node[circle, draw, inner sep=1pt] (v5) at (0,  2.8) {\tikz{\draw[] (0,0) ..controls (0.25,0) and (0.15,0.3) .. (0.4,0.3);}};
\node[circle, draw, inner sep=1pt] (v6) at (0,  3.6) {\tikz{\draw[] (0,0) ..controls (0.25,0) and (0.15,0.3) .. (0.4,0.3);}};
\end{scope}

\begin{scope}[xshift=6.0cm]
\node[circle, draw, inner sep=1pt] (u1) at (0,  0.4) {\tikz{\draw[] (0,0) ..controls (0.25,0) and (0.15,0.3) .. (0.4,0.3);}};
\node[circle, draw, inner sep=1pt] (u2) at (0,  1.6) {\tikz{\draw[] (0,0) ..controls (0.25,0) and (0.15,0.3) .. (0.4,0.3);}};
\node[circle, draw, inner sep=1pt] (u3) at (0,  2.8) {\tikz{\draw[] (0,0) ..controls (0.25,0) and (0.15,0.3) .. (0.4,0.3);}};
\end{scope}

\begin{scope}[xshift=7.0cm]
\node[inner sep=1pt] (o1) at (0,  0.4) {${y}_1$};
\node[inner sep=1pt] (o2) at (0,  1.6) {${y}_2$};
\node[inner sep=1pt] (o3) at (0,  2.8) {${y}_3$};
\end{scope}

\foreach \i in {1,...,5}{
\foreach \j in {1,...,6}{
\draw[-{Latex[length=2mm,width=1.2mm]}] (in\i) -- (w\j);}}

\foreach \i in {1,...,6}{
\foreach \j in {1,...,6}{
\draw[-{Latex[length=2mm,width=1.2mm]}] (w\i) -- (v\j);}}

\foreach \i in {1,...,6}{
\foreach \j in {1,...,3}{
\draw[-{Latex[length=2mm,width=1.2mm]}] (v\i) -- (u\j);}}

\foreach \i in {1,...,3}{
\draw[-{Latex[length=2mm,width=1.2mm]}] (u\i) -- (o\i);}

\path[fill=col1!50!white] (-0.65, 4) -- (1.25, 4) -- (1.25, 4.5) -- (-0.65, 4.5) -- cycle;
\path[fill=col2!50!white] (1.35, 4) -- (4.95, 4) -- (4.95, 4.5) -- (1.35, 4.5) -- cycle;
\path[fill=col3!50!white] (5.05, 4) -- (7.25, 4) -- (7.25, 4.5) -- (5.05, 4.5) -- cycle;

\node[] at (0.3,4.25) {\small input layer};
\node[] at (3.2,4.25) {\small hidden layers};
\node[] at (6.1,4.25) {\small readout layer};

\end{tikzpicture}
\caption{Example for a simple neuronal network with five input values and three output values. Each circle represents a neuron including weights, bias and activation function, arrows represent the flow of stimuli.}
\label{fig:deepnet}
\end{figure}

Neuronal networks with arbitrary complexity can be created by linking cells to each other. Cells working serially create a so called deep network \cite{cheng2016wide}, increasing the complexity substantially. An example for a deep network is shown in Figure~\ref{fig:deepnet}. The example consists of five input nodes, $x_1$ to $x_5$, which get processed by two hidden layers with six neurons each and a readout layer with three neurons. This network connects five input features (whatever they may be for a certain task) with three output features. The output is usually the probability that input features match a certain category. All neurons of a layer are connected to all neurons of the following layer. This architecture is called densely connected network.

A single densely connected layer can be described with the function
\begin{align}
\b{y} = f\left(\b{W}\,\b{x}+\b{b}\right),
\end{align}
where $\b{x}$ and $\b{y}$ are the vectors of neuron inputs and outputs, $\b{W}$ the weight tensor and $\b{b}$ the bias vector. The full network consists of nested layers and can thus create a very complex non-linear function or model, which can be expanded to
\begin{align}
\b{y} = f\left(\b{W}_3\,f\left(\b{W}_2\,f\left(\b{W}_1\,\b{x}+\b{b}_1\right)+\b{b}_2\right)+\b{b}_3\right).\label{eq:complete_net}
\end{align}
The first layer of the network in Figure~\ref{fig:deepnet} contains six neurons with five inputs (e.g.\ dendrites) per neuron. Therefore the mathematical description requires 30 weights and six biases for this layer. The whole network contains 84 weights and 15 biases, which can be summarised as network parameters $\bs\theta = \{\b{W}_i, \b{b}_{i}\}$. Network parameters represent the networks degree of freedom. A high degree of freedom means that the model can describe complex processes but also that learning is difficult. The models complexity can be easily adjusted due to the modular architecture.

\paragraph{Learning}

The result of the network depends on the parameters. A neuronal network is worthless without the appropriate parameters. Finding the optimal set of parameters for a specific network and a specific task is called learning or training. We stick to the simplest training method, called supervised learning \cite{schmidhuber2015deep}. To train a network we need three additional components: (1) training data which contains examples for in- and output pairs, (2) a cost or loss function, describing the fitness of the network to the training data and (3) an optimiser which modifies the parameters to minimise the cost function. 

The training data has to provide a set of in- and output pairs, $\b{x}'_k$ and $\b{y}'_k$. Based on the input vector $\b{x}'_k$, the network can calculate an answer $\b{y}_k= f_{\mathrm{net}}\left(\b{x}_k'\right)$. This step is called the forward pass. It is now the goal to minimise the difference between the correct answer $\b{y}'_k$ and the networks answer $\b{y}_k$ for all pairs $k$. The difference between vectors $\b{y}_k$ and $\b{y}_k'$ is defined by the cost function. There are many definitions for this function, the most common one is the so called cross entropy, defined as
\begin{align}
L_k = -\sum\limits_i y_{i,k}'\,\log\left(y_{i,k}\right)\label{eq:loss}
\end{align}
for a single data pair $\b{x}_k'$, $\b{y}_k'$. Finally, the optimiser has to find values for all network parameters $\bs\theta$, such that the mean loss $L = \frac{1}{K}\sum L_k$
for all training data pairs is minimised. The loss function~\eqref{eq:loss} and the network function~\eqref{eq:complete_net} are both steady and differentiable with respect to $\bs\theta$. For these special kind of functions, gradient descent optimiser are well suited. These methods work by differentiating the loss with respect to the parameters. The resulting gradient points towards the steepest growth of the loss. Therefore, stepping in the opposite direction of the gradient leads to a reduction of the loss and towards the optimum parameters. A step of the optimiser can be written as 
\begin{align}
\bs\theta^{i+1} = \bs\theta^{i}-\gamma\,\bs\nabla L,
\end{align}
where
\begin{align}
\bs\nabla L = \left(\dfrac{\partial\,L}{\partial \theta_1}, \dfrac{\partial\,L}{\partial \theta_2},\dots, \dfrac{\partial\,L}{\partial \theta_n}\right)^\r{T}
\end{align}
is the gradient of the loss $L$ with respect to the parameters $\bs\theta$, and $\gamma$ the learning rate which has to be chosen carefully by the developer. The gradient can be efficiently calculated by applying the chain rule starting from the top of the network. This step is called backward propagation. A complete learning step consists of a forward propagation yielding the loss, followed by a backward propagation yielding an improved parameter set. The classic gradient descent optimiser has been improved significantly in the last decade, leading to statistical gradient descent, momentum methods, AdaGrad, RMSProp and the current state of the art, the Adam method \cite{kingma2014adam}.

The optimisation of neuronal networks is an ill-posed problem, meaning that there is not a unique solution or that the solution changes drastically with minimal variation in the input. Latter is revealed by overfitting of the network to the training data. In this case, the network will not generalise to data it has not seen during training. This major problem has been solved in the last decade with powerful and efficient regularisation methods, such as dropout \cite{srivastava2014dropout}. However, also the best methods have limits, since there has to be sufficient data to identify statistically significant coherence. Usually a part of the available training data is not used for training and reserved for validation of the final network. This way one can identify overfitting and estimate a realistic performance.

\paragraph{Convolutional Neuronal Networks} 
The rapid progress in recent years can be attributed to the availability of cheap and fast hardware and the development of specialised network architectures, especially convolutional neuronal networks. In this context specialisation means to create copies of neurons and use them multiple times in a specific layout to share weights. This increases the complexity of the network without increasing its degree of freedom and thus learning difficulties. Convolutional neuronal networks (CNN) are inspired by the visual cortex of mammalians. These networks consider spatial relations and may also include spatial invariance. This makes them well suited for spatial problems such as the two major tasks of computer vision, image recognition and segmentation. Major attribution has to be given to Hubel and Wiesel \cite{hubel1959receptive} for their experiments on the visual cortex. They discovered the hierarchical structure of the visual cortex, consisting of simple, complex and hypercomplex cells with a limited receptive field per cell. Simple cells in the first hierarchical layer are able to recognise edges with different orientations in an image. Information about detected edges is further processed by the complex cells to recognise primitive shapes like circles and corners. This hierarchy continues and more complex objects are recognised in upper layers. 

\begin{figure}[htb]
\includegraphics[width=\textwidth]{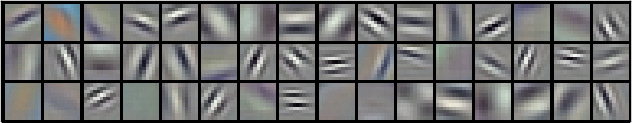}
\caption{Visualisation of some filters from the first layer of AlexNet \cite{krizhevsky2012imagenet}, which are scanning the underlying picture for edges with different orientations. This is clearly represented by the values in the filter, here visualised by colours. Note that these filters are learned by the network on its own and not preset.}
\label{fig:filter}
\end{figure}

There is a large amount of simple cells in a mammalian brain. Many of them have to identify the same edge but in a different area of the image. In other words, many cells are identical but have a different receptive field. In terms of the McCulloc-Pits model this means that many of the neurons share weights and the bias, but are linked to different input cells. This leads to a mathematical description similar to a convolution, thus the name convolutional neuronal network. The mathematical description of a simple convolutional layer as used for image recognition (two dimensional data) is
\begin{align}
y_{ij} = f\left(\sum\limits_{k=0}^{k<K} \sum\limits_{l=0}^{l<L} W_{kl}\,x_{i-(K-1)/2+k,j-(L-1)/2+l}+b\right),
\end{align}
where $K\times L$ is the filter size, $W_{kl}$ the weight matrix, in this context called filter, $b$ the bias. $y_{ij}$ is the result of the neuron at position $i,j$, $x_{ij}$ the input at position $i,j$. The result $y_{ij}$ of the convolutional layer is called feature map, since it describes the position where a specific feature can be found in the image. Although the mathematical description is complex, the concept is quite simple: one can imagine a convolutional layer to behave like a filter, moving over an image and emitting a stimulus when the filter matches an underlying structure. To match structures of different types, one has to make use of many such layers at a single hierarchical step, creating many feature maps per step. Going up in the hierarchy, the consecutive layers are finding structures in these three dimensional feature maps and output feature maps on their own. Examples for filters of the first layer, detecting edges of different orientation, are shown in Figure~\ref{fig:filter}.

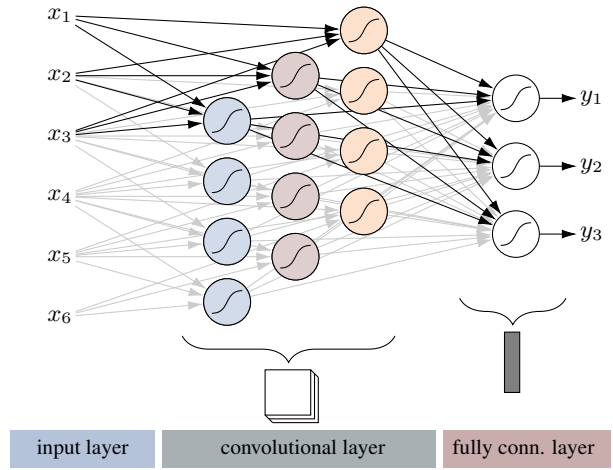
\begin{figure}[htb]
\begin{tikzpicture}[yscale=-1]

\node[inner sep=1pt] (in1) at (0,0) {$x_1$};
\node[inner sep=1pt] (in2) at (0,0.8) {$x_2$};
\node[inner sep=1pt] (in3) at (0,1.6) {$x_3$};
\node[inner sep=1pt] (in4) at (0,2.4) {$x_4$};
\node[inner sep=1pt] (in5) at (0,3.2) {$x_5$};
\node[inner sep=1pt] (in6) at (0,4.0) {$x_6$};

\begin{scope}[yshift=0.6cm,xshift=2.2cm]
\node[circle, draw, inner sep=1pt, fill=white] (w11) at (0,  0.8) {\tikz{\draw[] (0,0) ..controls (0.25,0) and (0.15,0.3) .. (0.4,0.3);}};
\node[circle, draw, inner sep=1pt, fill=white] (w12) at (0,  1.6) {\tikz{\draw[] (0,0) ..controls (0.25,0) and (0.15,0.3) .. (0.4,0.3);}};
\node[circle, draw, inner sep=1pt, fill=white] (w13) at (0,  2.4) {\tikz{\draw[] (0,0) ..controls (0.25,0) and (0.15,0.3) .. (0.4,0.3);}};
\node[circle, draw, inner sep=1pt, fill=white] (w14) at (0,  3.2) {\tikz{\draw[] (0,0) ..controls (0.25,0) and (0.15,0.3) .. (0.4,0.3);}};
\end{scope}

\begin{scope}[yshift=0cm, xshift=3.1cm]
\node[circle, draw, inner sep=1pt, fill=white] (w21) at (0,  0.8) {\tikz{\draw[] (0,0) ..controls (0.25,0) and (0.15,0.3) .. (0.4,0.3);}};
\node[circle, draw, inner sep=1pt, fill=white] (w22) at (0,  1.6) {\tikz{\draw[] (0,0) ..controls (0.25,0) and (0.15,0.3) .. (0.4,0.3);}};
\node[circle, draw, inner sep=1pt, fill=white] (w23) at (0,  2.4) {\tikz{\draw[] (0,0) ..controls (0.25,0) and (0.15,0.3) .. (0.4,0.3);}};
\node[circle, draw, inner sep=1pt, fill=white] (w24) at (0,  3.2) {\tikz{\draw[] (0,0) ..controls (0.25,0) and (0.15,0.3) .. (0.4,0.3);}};
\end{scope}

\begin{scope}[yshift=-0.6cm, xshift=4.0cm]
\node[circle, draw, inner sep=1pt, fill=white] (w31) at (0,  0.8) {\tikz{\draw[] (0,0) ..controls (0.25,0) and (0.15,0.3) .. (0.4,0.3);}};
\node[circle, draw, inner sep=1pt, fill=white, fill=white] (w32) at (0,  1.6) {\tikz{\draw[] (0,0) ..controls (0.25,0) and (0.15,0.3) .. (0.4,0.3);}};
\node[circle, draw, inner sep=1pt, fill=white] (w33) at (0,  2.4) {\tikz{\draw[] (0,0) ..controls (0.25,0) and (0.15,0.3) .. (0.4,0.3);}};
\node[circle, draw, inner sep=1pt, fill=white] (w34) at (0,  3.2) {\tikz{\draw[] (0,0) ..controls (0.25,0) and (0.15,0.3) .. (0.4,0.3);}};
\end{scope}

\begin{scope}[xshift=6.0cm]
\node[circle, draw, inner sep=1pt] (u1) at (0,  1.1) {\tikz{\draw[] (0,0) ..controls (0.25,0) and (0.15,0.3) .. (0.4,0.3);}};
\node[circle, draw, inner sep=1pt] (u2) at (0,  2.0) {\tikz{\draw[] (0,0) ..controls (0.25,0) and (0.15,0.3) .. (0.4,0.3);}};
\node[circle, draw, inner sep=1pt] (u3) at (0,  2.9) {\tikz{\draw[] (0,0) ..controls (0.25,0) and (0.15,0.3) .. (0.4,0.3);}};
\end{scope}

\begin{scope}[xshift=7.0cm]
\node[inner sep=1pt] (o1) at (0,  1.1) {${y}_1$};
\node[inner sep=1pt] (o2) at (0,  2.0) {${y}_2$};
\node[inner sep=1pt] (o3) at (0,  2.9) {${y}_3$};
\end{scope}

\foreach \i in {2,3,4}{
\foreach \j in {\number\numexpr\i,...,\number\numexpr\i+2}{
\draw[black!20!white,-{Latex[length=2mm,width=1.2mm]}] (in\j) -- (w1\i);}}
\foreach \i in {2,3,4}{
\foreach \j in {\number\numexpr\i,...,\number\numexpr\i+2}{
\draw[black!20!white,-{Latex[length=2mm,width=1.2mm]}] (in\j) -- (w2\i);}}
\foreach \i in {2,3,4}{
\foreach \j in {\number\numexpr\i,...,\number\numexpr\i+2}{
\draw[black!20!white,-{Latex[length=2mm,width=1.2mm]}] (in\j) -- (w3\i);}}

\foreach \i in {2,...,4}{
\foreach \j in {1,...,3}{
\draw[black!20!white, -{Latex[length=2mm,width=1.2mm]}] (w1\i) -- (u\j);}}
\foreach \i in {2,...,4}{
\foreach \j in {1,...,3}{
\draw[black!20!white, -{Latex[length=2mm,width=1.2mm]}] (w2\i) -- (u\j);}}
\foreach \i in {2,...,4}{
\foreach \j in {1,...,3}{
\draw[black!20!white, -{Latex[length=2mm,width=1.2mm]}] (w3\i) -- (u\j);}}

\foreach \i in {1}{
\foreach \j in {\number\numexpr\i,...,\number\numexpr\i+2}{
\draw[-{Latex[length=2mm,width=1.2mm]}] (in\j) -- (w1\i);}}
\foreach \i in {1}{
\foreach \j in {\number\numexpr\i,...,\number\numexpr\i+2}{
\draw[-{Latex[length=2mm,width=1.2mm]}] (in\j) -- (w2\i);}}
\foreach \i in {1}{
\foreach \j in {\number\numexpr\i,...,\number\numexpr\i+2}{
\draw[-{Latex[length=2mm,width=1.2mm]}] (in\j) -- (w3\i);}}

\foreach \i in {1}{
\foreach \j in {1,...,3}{
\draw[-{Latex[length=2mm,width=1.2mm]}] (w1\i) -- (u\j);}}
\foreach \i in {1}{
\foreach \j in {1,...,3}{
\draw[-{Latex[length=2mm,width=1.2mm]}] (w2\i) -- (u\j);}}
\foreach \i in {1}{
\foreach \j in {1,...,3}{
\draw[-{Latex[length=2mm,width=1.2mm]}] (w3\i) -- (u\j);}}

\foreach \i in {1,...,3}{
\draw[-{Latex[length=2mm,width=1.2mm]}] (u\i) -- (o\i);}

\begin{scope}[yshift=0.6cm,xshift=2.2cm]
\node[circle, draw, inner sep=1pt, fill=col1!30!white] (w11) at (0,  0.8) {\tikz{\draw[] (0,0) ..controls (0.25,0) and (0.15,0.3) .. (0.4,0.3);}};
\node[circle, draw, inner sep=1pt, fill=col1!30!white] (w12) at (0,  1.6) {\tikz{\draw[] (0,0) ..controls (0.25,0) and (0.15,0.3) .. (0.4,0.3);}};
\node[circle, draw, inner sep=1pt, fill=col1!30!white] (w13) at (0,  2.4) {\tikz{\draw[] (0,0) ..controls (0.25,0) and (0.15,0.3) .. (0.4,0.3);}};
\node[circle, draw, inner sep=1pt, fill=col1!30!white] (w14) at (0,  3.2) {\tikz{\draw[] (0,0) ..controls (0.25,0) and (0.15,0.3) .. (0.4,0.3);}};
\end{scope}

\begin{scope}[yshift=0cm, xshift=3.1cm]
\node[circle, draw, inner sep=1pt, fill=col3!30!white] (w21) at (0,  0.8) {\tikz{\draw[] (0,0) ..controls (0.25,0) and (0.15,0.3) .. (0.4,0.3);}};
\node[circle, draw, inner sep=1pt, fill=col3!30!white] (w22) at (0,  1.6) {\tikz{\draw[] (0,0) ..controls (0.25,0) and (0.15,0.3) .. (0.4,0.3);}};
\node[circle, draw, inner sep=1pt, fill=col3!30!white] (w23) at (0,  2.4) {\tikz{\draw[] (0,0) ..controls (0.25,0) and (0.15,0.3) .. (0.4,0.3);}};
\node[circle, draw, inner sep=1pt, fill=col3!30!white] (w24) at (0,  3.2) {\tikz{\draw[] (0,0) ..controls (0.25,0) and (0.15,0.3) .. (0.4,0.3);}};
\end{scope}

\begin{scope}[yshift=-0.6cm, xshift=4.0cm]
\node[circle, draw, inner sep=1pt, fill=col5!30!white] (w31) at (0,  0.8) {\tikz{\draw[] (0,0) ..controls (0.25,0) and (0.15,0.3) .. (0.4,0.3);}};
\node[circle, draw, inner sep=1pt, fill=col5!30!white] (w32) at (0,  1.6) {\tikz{\draw[] (0,0) ..controls (0.25,0) and (0.15,0.3) .. (0.4,0.3);}};
\node[circle, draw, inner sep=1pt, fill=col5!30!white] (w33) at (0,  2.4) {\tikz{\draw[] (0,0) ..controls (0.25,0) and (0.15,0.3) .. (0.4,0.3);}};
\node[circle, draw, inner sep=1pt, fill=col5!30!white] (w34) at (0,  3.2) {\tikz{\draw[] (0,0) ..controls (0.25,0) and (0.15,0.3) .. (0.4,0.3);}};
\end{scope}

\draw [decorate,decoration={brace,amplitude=10pt},xshift=-4pt,yshift=0pt]
(4.55, 4.25) -- (1.75, 4.25) node [midway,yshift=18pt, rotate=90] (cc) {};

\begin{scope}[xshift=3.0cm, yshift=5cm]
\draw[] (-0.3, -0.3) -- (-0.3, 0.3) -- (0.3, 0.3) -- (0.3, -0.3) -- cycle;
\draw[] (-0.25, 0.35) -- (0.35, 0.35) -- (0.35, -0.25);
\draw[] (-0.2, 0.40) -- (0.40, 0.40) -- (0.4, -0.2);
\draw[] (-0.3, 0.3) -- (-0.20, 0.40);
\draw[] (0.3, -0.3) -- (0.40, -0.20);
\end{scope}

\draw [decorate,decoration={brace,amplitude=10pt},xshift=-4pt,yshift=0pt]
(6.8, 3.75) -- (5.4, 3.75) node [midway,yshift=18pt, rotate=90] (cc) {};

\begin{scope}[xshift=5.95cm, yshift=4.6cm]
\draw[fill=black!50!white] (-0.1, -0.4) -- (-0.1, 0.4) -- (0.1, 0.4) -- (0.1, -0.4) -- cycle;
\end{scope}

\path[fill=col1!50!white] (-0.65, 5.5) -- (1.25, 5.5) -- (1.25, 6) -- (-0.65, 6) -- cycle;
\path[fill=col2!50!white] (1.35, 5.5) -- (4.95, 5.5) -- (4.95, 6) -- (1.35, 6) -- cycle;
\path[fill=col3!50!white] (5.05, 5.5) -- (7.25, 5.5) -- (7.25, 6) -- (5.05, 6) -- cycle;

\node[] at (0.3,5.75) {\small input layer};
\node[] at (3.2,5.75) {\small convolutional layer};
\node[] at (6.1,5.75) {\small fully conn. layer};

\end{tikzpicture}
\caption{Example for a simple one dimensional convolutional network with a single hidden convolutional layer with depth three. Coloured neurons share their parameters $\bs\theta$. Although we have the same amount of complexity as before, the number of parameters is reduced. Connections to and from neurons at position one are drawn black, all other in grey to reduce confusion.}
\label{fig:convnet}
\end{figure}

Convolutional layers are often followed up by pooling layers to increase the information density. These layers simply forward the maximum stimulus of a group of neurons. Finally on the very top, there are one or more fully connected layers, as known from simple neuronal networks, which output the probability for the image to match a specific category. To output plausible probabilities a softmax function is applied on top of the last layer. A simple one dimensional convolutional neuronal network is shown in Figure~\ref{fig:convnet}. It consists of the same amount of neurons as the fully connected network in Figure~\ref{fig:deepnet} but contains only half its parameters (45 weights and six biases), most of them in the last fully connected layer.

A convolutional network was first used by LeCun et al.\ in 1998 \cite{lecun1998gradient} to classify hand written digits, reaching an error rate below 1\%.
Until the year 2014 nobody thought that this approach could be transferred to large scale image recognition. The decisive breakthrough has been achieved by Krizhevsky et al.\ in 2012 \cite{krizhevsky2012imagenet} by using two GPUs to run a convolutional neuronal network with 60 million parameters. They achieved a top-1 error rate of 37.5\% and top-5 error rate of 17.0\% (the correct prediction is under the five outputs with the highest probability) in the yearly imagenet competition. This was groundbreaking at that time. All following competitions have been won by evolutions of Krizhevskys network (called AlexNet). 
State of the art CNNs consist of more than 100 convolutional layers with up to 512 feature maps per layer and millions of parameters. These networks are trained on multiple high performance GPUs for a few weeks with millions of training samples and reach top-5 error rates below 4\% \cite{he2016deep}.


\section{A Neuronal Network for Hazard Zone Mapping}

The task we chose to try in this work is strongly related to semantic segmentation, a task where the network has to mark position and form of objects in images. In fact, semantic segmentation gave us the inspiration for this work. Instead of marking positions of recognised objects in images, we mark vulnerable areas on a map.  

\begin{figure}[htb]
\includegraphics[width=0.45\textwidth]{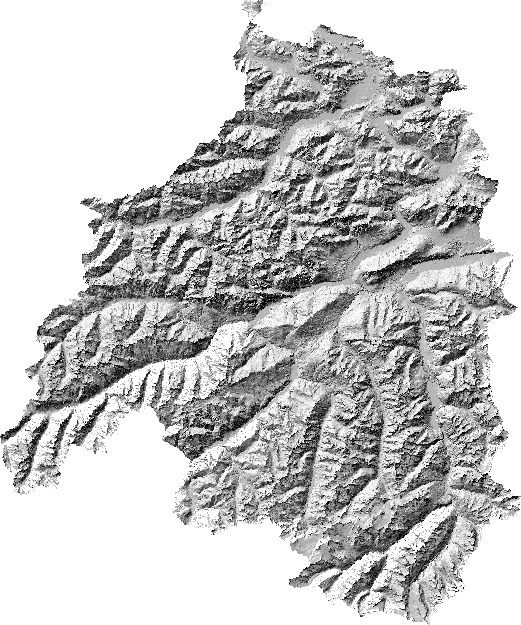}
\includegraphics[width=0.45\textwidth]{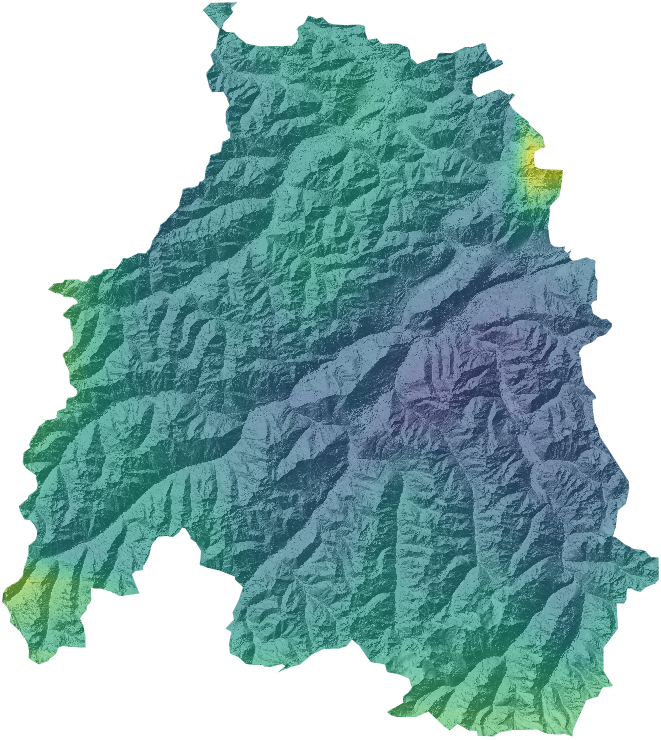}
\caption{The input data for the neuronal network: The elevation map shown as hill shade (left) and the snow height map (right, hill shade in background) of the Tirolean Oberland.}
\label{fig:input}
\end{figure}

\begin{figure}[h!]
\begin{tikzpicture}
\node[] () at (0,0) {\includegraphics[width=0.5\textwidth]{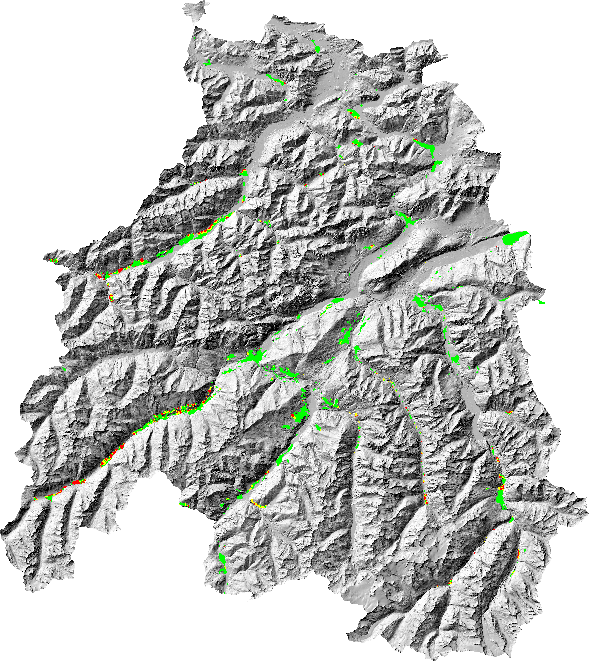}};

\node[inner sep = 0pt] (b) at (-4.5,0.5) {\includegraphics[width=0.4\textwidth]{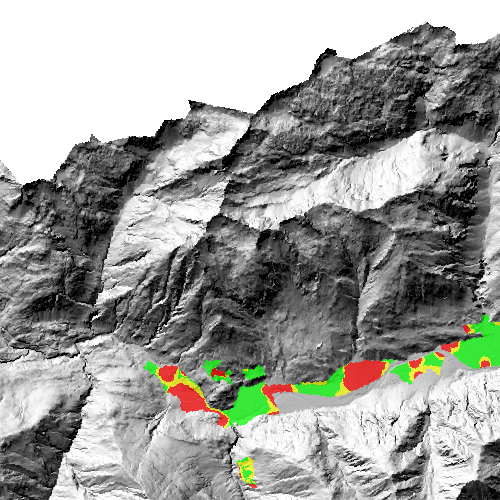}};
\node[draw, inner sep = 7.pt] (s) at (-1.45, 0.6) {};

\path[draw] (b.north east) -- (b.north west) -- (b.south west) -- (b.south east) -- cycle;
\path[clip] (b.north east) -- (b.south east) -- +(5, 0) -- +(0,5)-- cycle;
\draw[] (b.north east) -- (s.north east);
\draw[] (b.south east) -- (s.south east);
\draw[] (b.north west) -- (s.north west);
\draw[] (b.south west) -- (s.south west);
\end{tikzpicture}
\caption{The training data for the neuronal network: The hazard zone map of the Tirolean Oberland. Two regions (Längenfeld and Stanzer Valley) have been excluded from the training set and used for validation.}
\label{fig:trainingdata}
\end{figure}
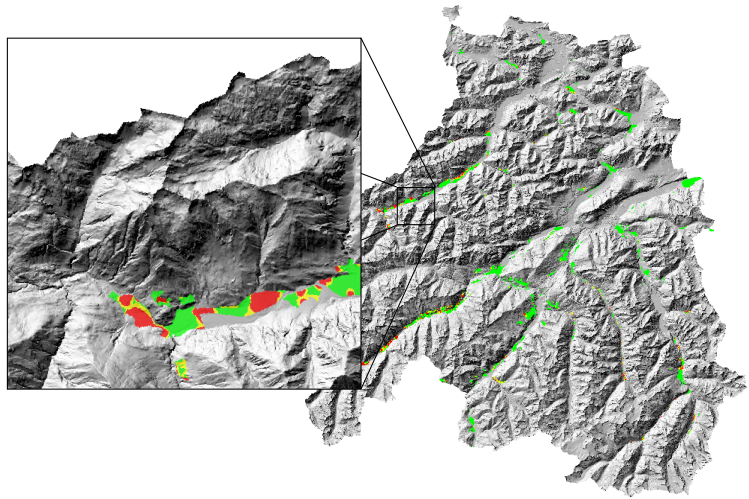

\begin{figure}[h!]
\centering
\begin{tikzpicture}
\node[inner sep=0pt] (m0) at (0,-0.05) {\includegraphics[width=4cm]{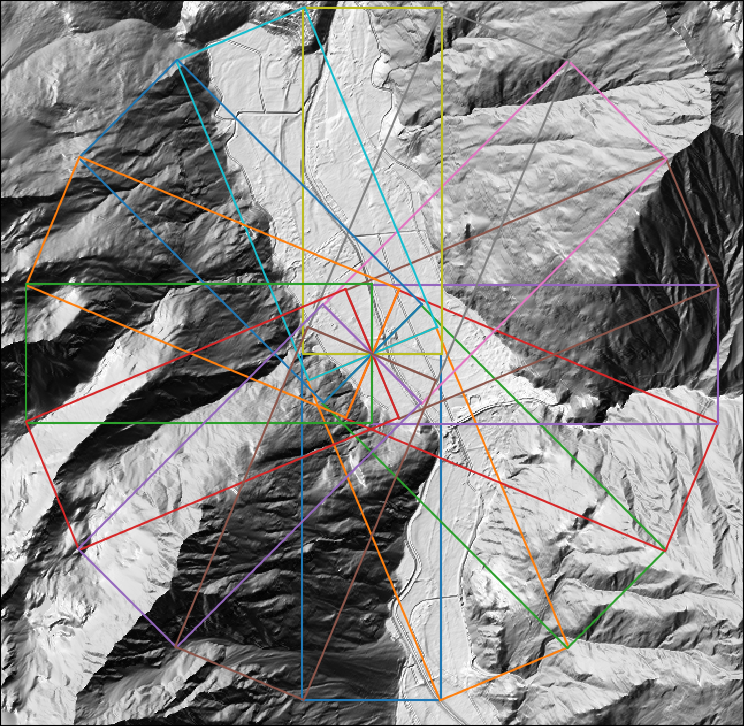}};
\begin{scope}[rotate=180]
\path[fill=col1, opacity=0.5] (-0.375, 0) -- (0.375,0) -- (0.375, 1.85) -- (-0.375, 1.85) -- cycle;
\draw[col1, thick] (-0.375, 0) -- (0.375,0) -- (0.375, 1.85) -- (-0.375, 1.85) -- cycle;
\node[inner sep=0pt] (anc) at (0, 1.85) {};
\end{scope}
\draw[col1] (anc) -- +(0,-0.5) node[below,black]{\tiny viewport 1};

\begin{scope}[rotate=202.5]
\path[fill=col3, opacity=0.5] (-0.375, 0) -- (0.375,0) -- (0.375, 1.85) -- (-0.375, 1.85) -- cycle;
\draw[col3, thick] (-0.375, 0) -- (0.375,0) -- (0.375, 1.85) -- (-0.375, 1.85) -- cycle;
\node[inner sep=0pt] (anc) at (0, 1.85) {};
\end{scope}
\draw[col3] (anc) -- +(0,-0.45) node[below,black]{\tiny viewport 2};

\begin{scope}[rotate=225]
\path[fill=col5, opacity=0.5] (-0.375, 0) -- (0.375,0) -- (0.375, 1.85) -- (-0.375, 1.85) -- cycle;
\draw[col5, thick] (-0.375, 0) -- (0.375,0) -- (0.375, 1.85) -- (-0.375, 1.85) -- cycle;\node[inner sep=0pt] (anc) at (0, 1.85) {};
\end{scope}
\draw[col5] (anc) -- +(0,-0.65) node[below,black]{\tiny viewport 3};

\node[yscale=-1, inner sep=0pt] (c2) at (4,-1.4) {\includegraphics[angle=90, width=2.5cm]{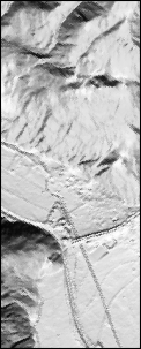}};
\node[yscale=-1, inner sep=0pt] (c1) at (4, -0.05) {\includegraphics[angle=90, width=2.5cm]{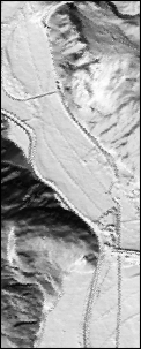}};
\node[yscale=-1, inner sep=0pt] (c0) at (4, 1.3) {\includegraphics[angle=90, width=2.5cm]{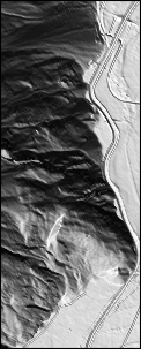}};

\draw[col1, line width = 1.5pt] (c0.north east) -- (c0.north west) -- (c0.south west) -- (c0.south east) -- cycle;
\draw[col3, line width = 1.5pt] (c1.north east) -- (c1.north west) -- (c1.south west) -- (c1.south east) -- cycle;
\draw[col5, line width = 1.5pt] (c2.north east) -- (c2.north west) -- (c2.south west) -- (c2.south east) -- cycle;

\node[xshift=-0.2cm, rotate=90] at (m0.west) {terrain map};
\node[xshift=-0.2cm, rotate=90] at (c0.west) {\tiny viewport 1};
\node[xshift=-0.2cm, rotate=90] at (c1.west) {\tiny viewport 2};
\node[xshift=-0.2cm, rotate=90] at (c2.west) {\tiny viewport 3};

\node[fill=black, circle, draw, inner sep=1.5pt] at (0,0) {};
\node[fill=black, circle, draw, inner sep=1.5pt] at (c0.east) {};
\node[fill=black, circle, draw, inner sep=1.5pt] at (c1.east) {};
\node[fill=black, circle, draw, inner sep=1.5pt] at (c2.east) {};

\draw [decorate,decoration={brace,amplitude=10pt},xshift=-4pt,yshift=0pt]
(5.5, 2) -- (5.5, -2) node [midway,xshift=18pt, rotate=90] {feed to network};

\end{tikzpicture}
\caption{To create a rotation invariant neuronal network we extract viewports forming a circle around the point of interest (black circle). The extracted patches are then feed into different sub-networks. Here the extraction of the first three viewports is shown. A similar procedure is done with the snow map but with a down scaled resolution.}
\label{fig:viewport}
\end{figure}

As indicators or input data for the vulnerability of a point we chose two geographic maps surrounding this point: (1) the terrain (Figure~\ref{fig:input}, left) and (2) the maximum expected snowfall within a period of three days (Figure~\ref{fig:input}, right). Although there are more indicators for avalanches, we identified these as the most important ones and sufficient for a proof of concept.

An indicator we eliminated was the exposition of the slope. This may seem irritating for an experienced alpinist. However, for catastrophic avalanches, which usually happen after a long time without sunshine, the exposition does not matter. We used this fact to apply a rotation invariant convolutional neuronal network. This idea has already been successfully employed by Dieleman et al.\ \cite{dieleman2015rotation} to classify galaxies. 

The desired output of our network is the hazard zone map, as shown in Figure~\ref{fig:trainingdata}. To train the network we have to repeatedly feed it with the input maps and enforce the network to yield the hazard zone map by adapting its parameters.

The rotation invariant neuronal network is assembled with a set of convolutional neuronal networks with different receptive fields or viewports. The viewports are arranged around the point of interest, each looking in a different direction, as shown in Figure~\ref{fig:viewport}. The results of all sub-networks are combined by a fully connected network to get the final result for the vulnerability of the investigated point.

\begin{figure*}[h!]
\rotatebox{90}{
\scalebox{1.0}{
\begin{tikzpicture}[yscale=-1]
\node[yscale=-1, inner sep=0pt] (c0) at (5,0) {\includegraphics[angle=90, width=3cm]{patch0.png}};
\node[fill=black, circle, draw, inner sep=1.5pt] at (c0.east) {};

\node[yscale=-1, inner sep=0pt] (c0) at (5,2) {\includegraphics[angle=90, width=3cm]{patch1.png}};
\node[fill=black, circle, draw, inner sep=1.5pt] at (c0.east) {};

\begin{scope}[xshift=5cm, yshift=3.5cm]
\node[circle, inner sep=0.5pt, fill=black] at (0, -0.2) {};
\node[circle, inner sep=0.5pt, fill=black] at (0, 0) {};
\node[circle, inner sep=0.5pt, fill=black] at (0, 0.2) {};
\end{scope}

\node[yscale=-1, inner sep=0pt] (c0) at (5,5) {\includegraphics[angle=90, width=3cm]{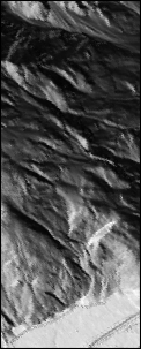}};
\node[fill=black, circle, draw, inner sep=1.5pt] at (c0.east) {};

\foreach \dy in {0, 2, 5}{
\begin{scope}[yshift=\dy cm]

	\begin{scope}[xshift=5cm, yshift=-0.1cm]
	\node[] (f1) at (2, 0.5) {};
	\node[] (f2) at (4, 0.5) {};
	\node[] (f3) at (4, -0.5) {};
	\node[] (f4) at (2, -0.5) {};

	\foreach \x in {0.05, 0.1, 0.15, 0.2}{
		\draw[] (f4.center) -- (f3.center) -- (f2.center) -- (f1.center) -- cycle;
		\path[] (f1.center) -- +(\x,\x) node (a1) {};
		\path[] (f2.center) -- +(\x,\x) node (a2) {};
		\path[] (f3.center) -- +(\x,\x) node (a3) {};
		\draw[] (a1.center) -- (a2.center) -- (a3.center);}
		\draw[] (f1.center) -- (a1.center);
		\draw[] (f2.center) -- (a2.center);
		\draw[] (f3.center) -- (a3.center);

	\node[] (c1) at (2.5, -0.1) {};
	\draw[col4] (c1.center) -- +(-3.2, -0.2) node (a1) {};
	\draw[col4] (c1.center) -- +(-2.8, -0.2) node (a2) {};
	\draw[col4] (c1.center) -- +(-2.8, 0.2) node (a3) {};
	\draw[col4] (c1.center) -- +(-3.2, 0.2) node (a4) {};

	\path[fill=col4, opacity=0.4] (a1.center) -- (a2.center) -- (a3.center) -- (a4.center); 
	\path[draw=col4] (a1.center) -- (a2.center) -- (a3.center) -- (a4.center) -- cycle; 
	\end{scope}

	\begin{scope}[xshift=7.5cm, yshift=-0.2cm]
	\node[] (f1) at (2, 0.35) {};
	\node[] (f2) at (3, 0.35) {};
	\node[] (f3) at (3, -0.35) {};
	\node[] (f4) at (2, -0.35) {};

	\foreach \x in {0.05, 0.1, 0.15,...,0.4}{
		\draw[] (f4.center) -- (f3.center) -- (f2.center) -- (f1.center) -- cycle;
		\path[] (f1.center) -- +(\x,\x) node (a1) {};
		\path[] (f2.center) -- +(\x,\x) node (a2) {};
		\path[] (f3.center) -- +(\x,\x) node (a3) {};
		\draw[] (a1.center) -- (a2.center) -- (a3.center);}
		\draw[] (f1.center) -- (a1.center);
		\draw[] (f2.center) -- (a2.center);
		\draw[] (f3.center) -- (a3.center);

	\node[] (c1) at (2.8, -0.2) {};
	\draw[col4] (c1.center) -- +(-1.95, -0.1) node (a1) {};
	\draw[col4] (c1.center) -- +(-1.55, -0.1) node (a2) {};
	\draw[col4] (c1.center) -- +(-1.55, 0.3) node (a3) {};
	\draw[col4] (c1.center) -- +(-1.95, 0.3) node (a4) {};

	\path[fill=col4, opacity=0.4] (a1.center) -- (a2.center) -- (a3.center) -- (a4.center); 
	\path[draw=col4] (a1.center) -- (a2.center) -- (a3.center) -- (a4.center) -- cycle; 
	\end{scope}

	\begin{scope}[xshift=9.2cm, yshift=-0.5cm]
	\node[] (f1) at (2, 0.25) {};
	\node[] (f2) at (2.7, 0.25) {};
	\node[] (f3) at (2.7, -0.25) {};
	\node[] (f4) at (2, -0.25) {};

	\foreach \x in {0.05, 0.1, 0.15,...,0.8}{
		\draw[] (f4.center) -- (f3.center) -- (f2.center) -- (f1.center) -- cycle;
		\path[] (f1.center) -- +(\x,\x) node (a1) {};
		\path[] (f2.center) -- +(\x,\x) node (a2) {};
		\path[] (f3.center) -- +(\x,\x) node (a3) {};
		\draw[] (a1.center) -- (a2.center) -- (a3.center);}
		\draw[] (f1.center) -- (a1.center);
		\draw[] (f2.center) -- (a2.center);
		\draw[] (f3.center) -- (a3.center);

	\node[] (c1) at (2.5, 0.1) {};
	\draw[col4] (c1.center) -- +(-1.35, 0.2) node (a1) {};
	\draw[col4] (c1.center) -- +(-1.55, 0.2) node (a2) {};
	\draw[col4] (c1.center) -- +(-1.55, 0.4) node (a3) {};
	\draw[col4] (c1.center) -- +(-1.35, 0.4) node (a4) {};

	\path[fill=col4, opacity=0.4] (a1.center) -- (a2.center) -- (a3.center) -- (a4.center); 
	\path[draw=col4] (a1.center) -- (a2.center) -- (a3.center) -- (a4.center) -- cycle; 
	\end{scope}

	\begin{scope}[xshift=10.5cm, yshift=-1cm]
	\node[] (f1) at (2, 0.10) {};
	\node[] (f2) at (2.3, 0.10) {};
	\node[] (f3) at (2.3, -0.10) {};
	\node[] (f4) at (2, -0.10) {};

	\foreach \x in {0.05, 0.1, 0.15,...,1.6}{
		\draw[] (f4.center) -- (f3.center) -- (f2.center) -- (f1.center) -- cycle;
		\path[] (f1.center) -- +(\x,\x) node (a1) {};
		\path[] (f2.center) -- +(\x,\x) node (a2) {};
		\path[] (f3.center) -- +(\x,\x) node (a3) {};
		\draw[] (a1.center) -- (a2.center) -- (a3.center);}
		\draw[] (f1.center) -- (a1.center);
		\draw[] (f2.center) -- (a2.center);
		\draw[] (f3.center) -- (a3.center);

	\node[] (c1) at (2.1, -0.05) {};
	\draw[col4] (c1.center) -- +(-1.15, 0.4) node (a1) {};
	\draw[col4] (c1.center) -- +(-1.35, 0.4) node (a2) {};
	\draw[col4] (c1.center) -- +(-1.35, 0.6) node (a3) {};
	\draw[col4] (c1.center) -- +(-1.15, 0.6) node (a4) {};

	\path[fill=col4, opacity=0.4] (a1.center) -- (a2.center) -- (a3.center) -- (a4.center); 
	\path[draw=col4] (a1.center) -- (a2.center) -- (a3.center) -- (a4.center) -- cycle; 
	\end{scope}

	\begin{scope}[xshift=13cm, yshift=-1.1cm]
	\node[] (f1) at (2, 0.10) {};
	\node[] (f2) at (2.2, 0.10) {};
	\path[] (f1.center) -- +(0, 1.5) node[] (f3){};
	\path[] (f2.center) -- +(0, 1.5) node[] (f4){};

	\draw[fill=black!50!white] (f1.center) -- (f2.center) -- (f4.center) -- (f3.center) -- cycle;

	\draw[col4] (f1.center) -- +(-2.2, -0.1) node (a1) {};
	\draw[col4] (f3.center) -- +(-0.6, 0.2) node (a2) {};
	\end{scope}

	\begin{scope}[xshift=13.75cm, yshift=-1.1cm]
	\node[] (f1) at (2, 0.10) {};
	\node[] (f2) at (2.2, 0.10) {};
	\path[] (f1.center) -- +(0, 1.5) node[] (f3){};
	\path[] (f2.center) -- +(0, 1.5) node[] (f4){};

	\draw[fill=black!50!white] (f1.center) -- (f2.center) -- (f4.center) -- (f3.center) -- cycle;

	\draw[col4] (f1.center) -- +(-0.8, 0) node (a1) {};
	\draw[col4] (f3.center) -- +(-0.8, 0) node (a2) {};
	\end{scope}

	\begin{scope}[xshift=14.5cm, yshift=-0.5cm]
	\node[] (f1) at (2, 0.10) {};
	\node[] (f2) at (2.2, 0.10) {};
	\path[] (f1.center) -- +(0, 0.3) node[] (f3){};
	\path[] (f2.center) -- +(0, 0.3) node[] (f4){};

	\draw[fill=black!50!white] (f1.center) -- (f2.center) -- (f4.center) -- (f3.center) -- cycle;

	\draw[col4] (f1.center) -- +(-0.55, -0.6) node (a1) {};
	\draw[col4] (f3.center) -- +(-0.55, 0.6) node (a2) {};
	\end{scope}
\end{scope}}

\begin{scope}[xshift=16cm, yshift=2.5cm]
	\node[] (f1) at (2, -1.6) {};
	\node[] (f2) at (2.2, -1.6) {};
	\path[] (f1.center) -- +(0, 3.2) node[] (f3){};
	\path[] (f2.center) -- +(0, 3.2) node[] (f4){};

	\draw[fill=black!50!white] (f1.center) -- (f2.center) -- (f4.center) -- (f3.center) -- cycle;

	\foreach \x in {-1.4,-1.2,...,1.5}{
		\draw[](2,\x) -- (2.2,\x);}

	\draw[col4] (2, -1.6) -- +(-1.3, -1.3) node (a1) {};
	\draw[col4] (2, -1.4) -- +(-1.3, -1.2) node (a1) {};

	\draw[col4] (2, -1.4) -- +(-1.3, 0.5) node (a1) {};
	\draw[col4] (2, -1.2) -- +(-1.3, 0.6) node (a1) {};

	\draw[col4] (2,  1.4) -- +(-1.3, 0.7) node (a1) {};
	\draw[col4] (2,  1.6) -- +(-1.3, 0.8) node (a1) {};
\end{scope}

\begin{scope}[xshift=16.75cm, yshift=2.5cm]

	\draw[fill=green] (2, -0.3) -- (2.2, -0.3) -- (2.2, -0.1) -- (2, -0.1) -- cycle;
	\draw[fill=yellow] (2, -0.1) -- (2.2, -0.1) -- (2.2, 0.1) -- (2, 0.1) -- cycle;
	\draw[fill=red] (2, 0.1) -- (2.2, 0.1) -- (2.2, 0.3) -- (2, 0.3) -- cycle;

	\draw[col4] (2, -0.3) -- +(-0.55, -1.3) node (a1) {};
	\draw[col4] (2,  0.3) -- +(-0.55, 1.3) node (a1) {};
\end{scope}

\node[circle, draw, fill=white, inner sep=1.5pt] (c0) at (5, 5.5) {};
\draw (c0) -- +(0, 0.5) node (c1) [below]{input maps};
\node[yshift=-0.1cm] at (c1.south) {$1\times84\times52$};

\node[circle, draw, fill=white, inner sep=1.5pt] (c0) at (8.2, 5.5) {};
\draw (c0) -- +(0, 0.5) node (c1) [below]{feature maps};
\node[yshift=-0.1cm] at (c1.south) {$8\times40\times24$};

\node[circle, draw, fill=white, inner sep=1.5pt] (c0) at (10.2, 5.4) {};
\draw (c0) -- +(0, 0.6) node (c1) [below]{feature maps};
\node[yshift=-0.1cm] at (c1.south) {$16\times18\times10$};

\node[circle, draw, fill=white, inner sep=1.5pt] (c0) at (12.2, 5.4) {};
\draw (c0) -- +(0, 1.4) node (c1) [below]{feature maps};
\node[yshift=-0.1cm] at (c1.south) {$(32+1)\times8\times4$};

\node[circle, draw, fill=white, inner sep=1.5pt] (c0) at (14.05, 5.5) {};
\draw (c0) -- +(0, 0.5) node (c1) [below]{feature maps};
\node[yshift=-0.1cm] at (c1.south) {$64\times3\times1$};

\node[circle, draw, fill=white, inner sep=1.5pt] (c0) at (15.1, 5.4) {};
\draw (c0) -- +(0, 1.4) node (c1) [below]{feature vector};
\node[yshift=-0.1cm] at (c1.south) {$512$};

\node[circle, draw, fill=white, inner sep=1.5pt] (c0) at (15.9, 5.4) {};
\draw (c0) -- +(0.5, 0.5) node (c1) [below]{feature vector};
\node[yshift=-0.1cm] at (c1.south) {$512$};

\node[circle, draw, fill=white, inner sep=1.5pt] (c0) at (16.6, 4.8) {};
\draw (c0) -- +(0.5, 0.5) node (c1) [below]{feature vector};
\node[yshift=-0.1cm] at (c1.south) {$3$};

\node[circle, draw, fill=white, inner sep=1.5pt] (c0) at (18.1, 4.0) {};
\draw (c0) -- +(0.5, 0.5) node (c1) [below]{joined feature vector};
\node[yshift=-0.1cm] at (c1.south) {$48$};

\node[circle, draw, fill=white, inner sep=1.5pt] (c0) at (18.85, 2.6) {};
\draw (c0) -- +(0.5, 0.5) node (c1) [below]{result vector};
\node[yshift=-0.1cm] at (c1.south) {$3$};

\node[circle, draw, fill=white, inner sep=1.5pt] (c0) at (6.8, -0.2) {};
\draw (c0) -- +(0, -0.65) node (c1) [above]{pool $2\times2$};
\node[yshift=0.1cm] at (c1.north) {conv $5\times5$};

\node[circle, draw, fill=white, inner sep=1.5pt] (c0) at (9.35, -0.35) {};
\draw (c0) -- +(-0.5, -0.5) node (c1) [above]{pool $2\times2$};
\node[yshift=0.1cm] at (c1.north) {conv $8\times5\times5$};

\node[circle, draw, fill=white, inner sep=1.5pt] (c0) at (11.0, -0.25) {};
\draw (c0) -- +(-0.5, -1.4) node (c1) [above]{pool $2\times2$};
\node[yshift=0.1cm] at (c1.north) {conv $16\times5\times5$};

\node[circle, draw, fill=white, inner sep=1.5pt] (c0) at (12.2, -0.9) {};
\draw (c0) -- +(0.5, -0.75) node (c1) [above]{pool $2\times2$};
\node[yshift=0.1cm] at (c1.north) {conv $33\times5\times5$};

\node[circle, draw, fill=white, inner sep=1.5pt] (c0) at (14.2, -0.7) {};
\draw (c0) -- +(0., -0.75) node (c1) [above]{dropout};
\node[yshift=0.1cm] at (c1.north) {fc1};
\node[circle, draw, fill=white, inner sep=1.5pt] (c0) at (15.5, -0.7) {};
\draw (c0) -- +(0, -0.75) node (c1) [above]{dropout};
\node[yshift=0.1cm] at (c1.north) {fc2};
\node[circle, draw, fill=white, inner sep=1.5pt] (c0) at (16.3, -0.4) {};
\draw (c0) -- +(0, -1.05) node (c1) [above]{fc3};

\node[circle, draw, fill=white, inner sep=1.5pt] (c0) at (17.4, 0.4) {};
\draw (c0) -- +(0, -1.05) node (c1) [above]{join operation};

\node[circle, draw, fill=white, inner sep=1.5pt] (c0) at (18.5, 2.2) {};
\draw (c0) -- +(0, -1.8) node (c1) [above]{read out};

\end{tikzpicture}
}
}
\caption{Architecture of the rotation invariant neuronal network. Numbers below feature maps and vectors show their dimensions. The snow map, which is not shown to reduce complexity, is appended to the feature maps of the third convolutional layer. The result vector contains the probabilities for the point of interest to be in the green, yellow or red zone. These probabilities are optimised to match the official hazard zone map during training.}
\label{fig:architecture}
\end{figure*}

Here we describe the architecture of the network which yields the best results. There is plenty of space for optimisation, which can be achieved by systematically changing so called meta-parameters, describing the architecture. We used 16 viewports with a physical dimension of $3360\,\mathrm{m}\times 2080\,\mathrm{m}$ and a resolution of $40\,\mathrm{m}/\mathrm{pixel}$, meaning that the network can ''look'' as far as $3360\,\mathrm{m}$ from the point under investigation. The viewports are processed by four convolutional layers with filter size $N\times5\times5$ or $N\times3\times3$, where $N$ is the number of feature maps the layer has to process. The first layer takes the viewports of the terrain map and outputs eight feature maps, the second layer takes those and outputs $16$ new feature maps, the third layer outputs $32$ and the fourth $64$. The snow map is attached to the feature maps of the third layer and passed to the fourth convolutional layer. We chose this approach because the information density in the snow map is much lower than in the terrain map. The fourth convolutional layer is followed up by two densely connected layers with $512$ neurons each and a fully connected layer with three neurons. Densely connected layers are followed up by dropout layers to reduce overfitting. The output of the last three neurons is the final result of the sub-network. The 16 sub-networks share all weights and are therefore completely identical. All sub-network results are merged and further processed by a final fully connected readout layer, yielding the probabilities of the three categories. The architecture is shown in Figure~\ref{fig:architecture}. Overall, the network contains approximately $400\,000$ parameters, most of them in the fully connected layers.

The neuronal network is implemented based on Googles machine learning toolkit TensorFlow \cite{abadi2016tensorflow}, one of the most efficient implementations for neuronal networks. The viewport extraction is running in parallel on the CPU, the processed viewports are continuously feed into the neuronal network, running on the GPU. This way we could utilise the available hardware efficiently.

\section{Training and Results}

The training data is shown in Figures~\ref{fig:input} (input, the $\b{x}'_k$ vectors) and \ref{fig:trainingdata} (desired output, the $\b{y}'_k$-vectors). The training data is unbalanced, meaning that the green zone is much bigger than red and yellow zones. In such a case, the network can achieve the best result by simply predicting the green zone all the time, which is an undesired behaviour. To address this issue we randomly subsample mini-batches of training data, where red, yellow and green zones are represented equally. 

\begin{figure}[htb]
\centering
\includegraphics[trim=0 1.2cm 10cm 0.3cm,clip, width=0.7\textwidth]{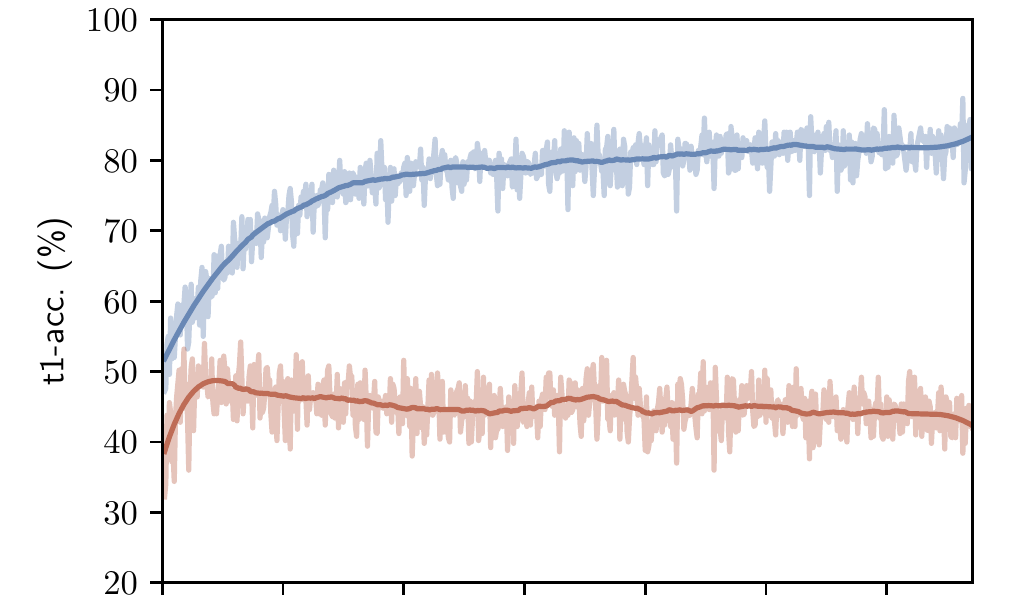}\\
\includegraphics[trim=10cm 0.3cm 0 0.3cm,clip, width=0.7\textwidth]{acc.pdf}
\caption{Top-1 (top) and top-2 accuracy (bottom) calculated on the balanced training set (blue) and the balanced validation set (red). The graphs are smoothed with a Savitzky-Gloay filter. The validation accuracy peaks after 2\,000 optimisation steps at approximately 50\%(95\%) accuracy. At this point training and validation diverge, which is a clear sign of overfitting.}
\label{fig:acc}
\end{figure}

All maps are randomly rotated and flipped to increase the entropy of the training data set. We used an Adam optimiser with an exponentially decaying learning rate. In the training phase we picked training and validation samples at random and continuously calculated the top-1 and top-2 accuracy. 
In here, the top-2 accuracy is defined as the probability for the network to yield the second best result (e.g.\ yellow instead of red). The top-1 (top-2) accuracy at start is 33\% (78\%), which corresponds to guessing. The training top-1 (top-2) accuracy is constantly raising up to approximately 85\% (99\%). We reach a peak top-1 (top-2) validation accuracy of approximately 50\% (95\%) on the balanced data set. Details of the accuracy evolution during the training phase are shown in Figure~\ref{fig:acc}. The large gap between training and validation accuracy is unusual for convolutional networks with dropout regularisation (c.f.\ Krizhevsky et al.\ \cite{krizhevsky2012imagenet}). We assume that this behaviour can be attributed to the small training data set which contains about 100 avalanches and 80\,000 data points per zone to determine 400\,000 parameters. The training has been stopped manually after 33\,000 steps which took $8.5\,\mathrm{h}$ on a machine with a Nvidia GTX 1060 GPU and an Intel i7-7000K CPU.

After the training phase, the network can be used to generate hazard zone maps for arbitrary regions, when the terrain and the expected snow cover is available. Automatic generated hazard zone maps for the validation regions are shown in Figures{to~\ref{fig:pred_lf} and \ref{fig:pred_st} alongside with the official hazard zone maps. We reduced predictions to areas where an official hazard zone map is available because only here we can judge results. These maps have been created with the final version of the network. We expect better results with the network at its peak accuracy. The generation of theses maps takes a few minutes on the machine used for training. We used two validation regions with substantially different topographies. The top-1 (top-2) accuracy of the generated hazard zone maps is approximately 85\% (94\%) for Längenfeld and 60\% (90\%) for the Stanzer Valley. These values are calculated on an unbalanced data set and therefore considerably higher. The reason is that green zones, which are easier to predict, are dominating in a real case scenario, especially in Längenfeld. 

The generated hazard zone maps show some interesting details. Looking at Längenfeld (Figure~\ref{fig:pred_lf}), one can observe that most avalanches have been identified by the neuronal network. However, the shape of the avalanches and their runout is represented poorly. Looking at the Stanzer Valley, one can see that green zones are preserved quite well, also in steep terrain. 

\begin{figure}[htb]
\includegraphics[width=0.9\textwidth]{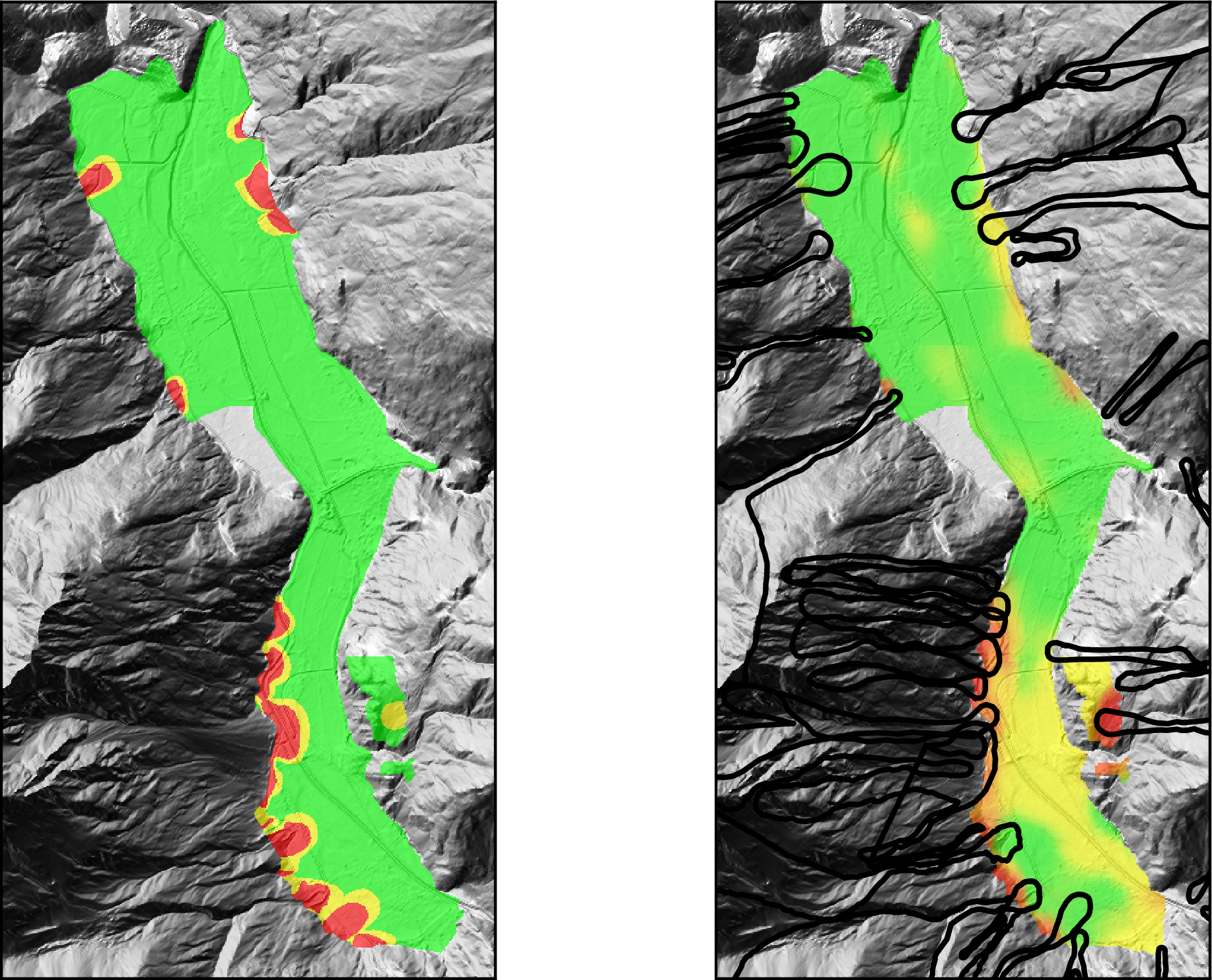}
\caption{The hazard zone map of Längenfeld (left) alongside with the prediction of the neuronal network (right). Black lines in the right picture show the outlines of documented avalanches. Note that the network did never see the official hazard zone map nor the documented avalanches.}
\label{fig:pred_lf}
\end{figure}

\begin{figure}[htb]
\includegraphics[width=0.9\textwidth]{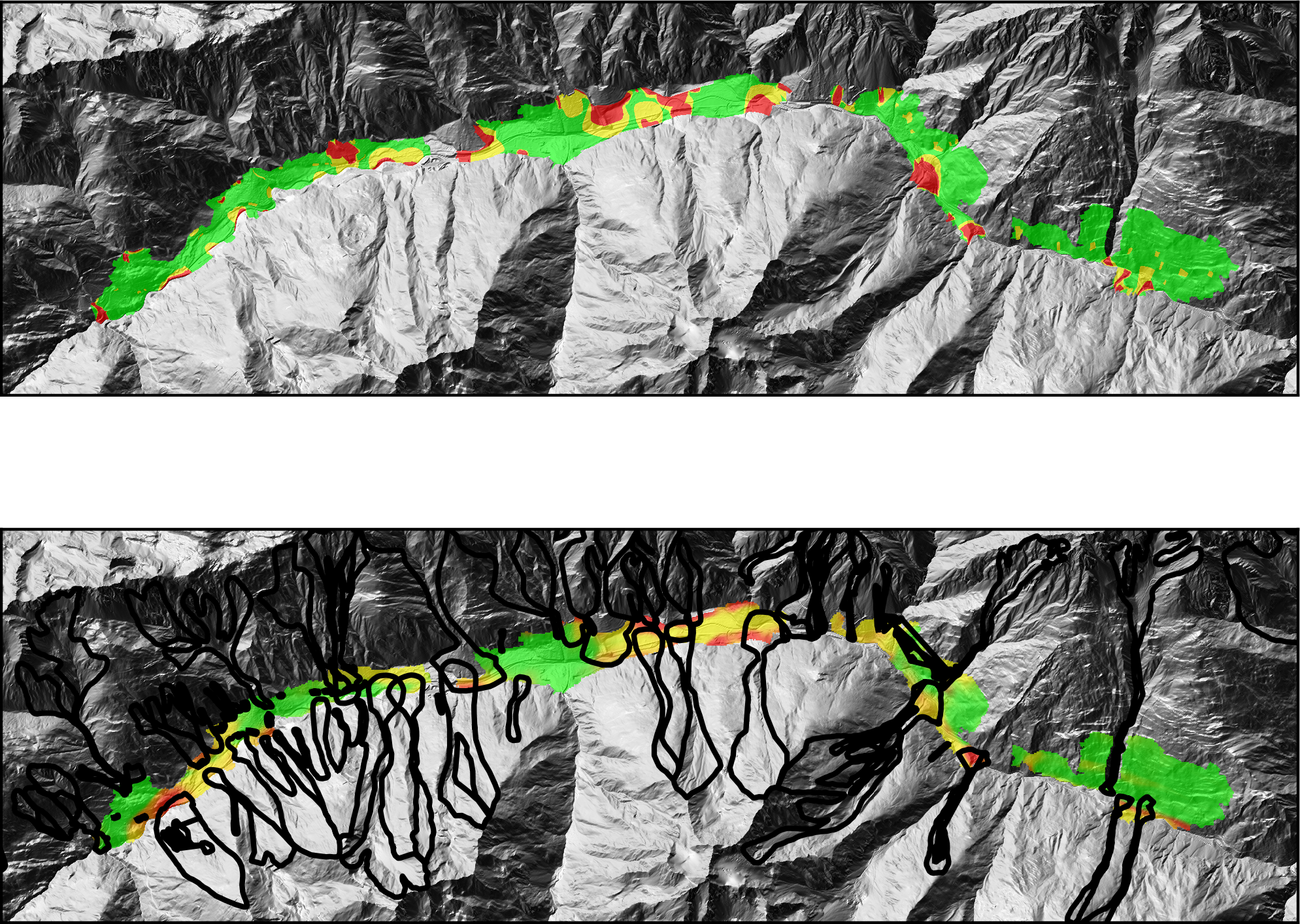}
\caption{The hazard zone map of Stanzer Valley (top) alongside with the prediction of the neuronal network (bottom). Black lines in the bottom picture show the outlines of documented avalanches. Note that the network did never see the official hazard zone map nor the documented avalanches.}
\label{fig:pred_st}
\end{figure}

\section{Summary and Outlook}

This work shows the application of a rotation invariant convolutional network to hazard zone mapping. The network learns from existing hazard zone maps and can generate new ones for other regions. This process allows us to transfer knowledge from one region to another.

The accuracy for two validation regions is significantly above plain coincidence, showing the feasibility of the idea. However, for a serious application, the approach needs to be further improved. There are several methods suggested in literature to improve the results, such as batch normalisation \cite{ioffe2015batch} or untied biases \cite{dieleman2015rotation}. Also the pooling of feature maps may be inappropriate for this task, because the related translation invariance may corrupt the estimation of runout distances. Overfitting of the network in an early training phase highlights the necessity for more training data. The demand for high amounts of training data is probably the biggest disadvantage of this method. However, there are training methods to improve results with small data sets, such as unsupervised learning \cite{schmidhuber2015deep}.

It is imperative to understand processes within the neuronal network before its practical application. There are several techniques to visualise the neuronal network and features which are determining its outcome \cite{zeiler2014visualizing}. This is not only interesting from a machine learning point of view, but may also lead to some new insights on avalanche formation since the neuronal network encodes statistical information of the training data.

Moreover, an important step will be the comparison with current methods. This is a difficult task, because most methods are not applicable on a regional scale and without human interaction.

Overall, we are confident that artificial intelligence and neuronal networks will play an important role in many areas of our everyday life. 
They have the potential to augment the creation process and the interregional standardization of hazard zone maps, which are a crucial instrument for safety considerations of settlement development, tourism and transit in alpine regions.

\section*{Acknowledgment}

{\footnotesize
We thank Prof.\ M.\ Harders and Prof.\ M.\ Haltmeier for supervision of this project and the Research Cluster \textit{Deep Learning} of the doctoral programme \textit{Computational Interdisciplinary Modelling} for valuable discussions. We gratefully acknowledge the support from M.\ Granig and F.\ \"Osterle (WLV) and from our doctoral advisors, Prof.\ W.\ Fellin and Prof.\ W.\ Rauch.}

{\footnotesize
\bibliographystyle{unsrt}
\bibliography{lit.bib}}

\begin{thebibliography}{10}

\bibitem{BMLFUW2011}
die.wildbach.
\newblock {Richtlinie für die Gefahrenzonenplanung}.
\newblock (BMLFUW-LE.3.3.3/0185-IV/5/2007), 2011.

\bibitem{BZ2012}
Landesregierung Autonome Provinz~Bozen Südtirol.
\newblock {Richtlinien zur Erstellung der Gefahrenzonenpläne (GZP) und zur
  Klassifizierung des spezifischen Risikos (KSR)}.
\newblock (Beiblatt Nr. 1 zum Amtsblatt Nr. 21/I-II vom 22/05/2012), 2012.

\bibitem{schmidhuber2015deep}
J{\"u}rgen Schmidhuber.
\newblock Deep learning in neural networks: An overview.
\newblock {\em Neural networks}, 61:85--117, 2015.

\bibitem{lied1980empirical}
Karstein Lied and Steinar Bakkehoi.
\newblock Empirical calculations of snow--avalanche run--out distance based on
  topographic parameters.
\newblock {\em Journal of Glaciology}, 26(94):165--177, 1980.

\bibitem{lee2004determination}
Saro Lee, Joo-Hyung Ryu, Joong-Sun Won, and Hyuck-Jin Park.
\newblock Determination and application of the weights for landslide
  susceptibility mapping using an artificial neural network.
\newblock {\em Engineering Geology}, 71(3):289--302, 2004.

\bibitem{isola2016image}
Phillip Isola, Jun-Yan Zhu, Tinghui Zhou, and Alexei~A Efros.
\newblock Image-to-image translation with conditional adversarial networks.
\newblock {\em arXiv preprint arXiv:1611.07004}, 2016.

\bibitem{mcculloch1943logical}
Warren~S McCulloch and Walter Pitts.
\newblock A logical calculus of the ideas immanent in nervous activity.
\newblock {\em The bulletin of mathematical biophysics}, 5(4):115--133, 1943.

\bibitem{cheng2016wide}
Heng-Tze Cheng, Levent Koc, Jeremiah Harmsen, Tal Shaked, Tushar Chandra,
  Hrishi Aradhye, Glen Anderson, Greg Corrado, Wei Chai, Mustafa Ispir, et~al.
\newblock Wide \& deep learning for recommender systems.
\newblock In {\em Proceedings of the 1st Workshop on Deep Learning for
  Recommender Systems}, pages 7--10. ACM, 2016.

\bibitem{kingma2014adam}
Diederik Kingma and Jimmy Ba.
\newblock Adam: A method for stochastic optimization.
\newblock {\em arXiv preprint arXiv:1412.6980}, 2014.

\bibitem{srivastava2014dropout}
Nitish Srivastava, Geoffrey~E Hinton, Alex Krizhevsky, Ilya Sutskever, and
  Ruslan Salakhutdinov.
\newblock Dropout: a simple way to prevent neural networks from overfitting.
\newblock {\em Journal of Machine Learning Research}, 15(1):1929--1958, 2014.

\bibitem{hubel1959receptive}
David~H Hubel and Torsten~N Wiesel.
\newblock Receptive fields of single neurones in the cat's striate cortex.
\newblock {\em The Journal of physiology}, 148(3):574--591, 1959.

\bibitem{krizhevsky2012imagenet}
Alex Krizhevsky, Ilya Sutskever, and Geoffrey~E Hinton.
\newblock Imagenet classification with deep convolutional neural networks.
\newblock In F.~Pereira, C.~J.~C. Burges, L.~Bottou, and K.~Q. Weinberger,
  editors, {\em Advances in Neural Information Processing Systems 25}, pages
  1097--1105. Curran Associates, Inc., 2012.

\bibitem{lecun1998gradient}
Yann LeCun, L{\'e}on Bottou, Yoshua Bengio, and Patrick Haffner.
\newblock Gradient-based learning applied to document recognition.
\newblock {\em Proceedings of the IEEE}, 86(11):2278--2324, 1998.

\bibitem{he2016deep}
Kaiming He, Xiangyu Zhang, Shaoqing Ren, and Jian Sun.
\newblock Deep residual learning for image recognition.
\newblock In {\em Proceedings of the IEEE Conference on Computer Vision and
  Pattern Recognition}, pages 770--778, 2016.

\bibitem{dieleman2015rotation}
Sander Dieleman, Kyle~W Willett, and Joni Dambre.
\newblock Rotation-invariant convolutional neural networks for galaxy
  morphology prediction.
\newblock {\em Monthly notices of the royal astronomical society},
  450(2):1441--1459, 2015.

\bibitem{abadi2016tensorflow}
Mart{\'\i}n Abadi, Ashish Agarwal, Paul Barham, Eugene Brevdo, Zhifeng Chen,
  Craig Citro, Greg~S Corrado, Andy Davis, Jeffrey Dean, Matthieu Devin, et~al.
\newblock Tensorflow: Large-scale machine learning on heterogeneous distributed
  systems.
\newblock {\em arXiv preprint arXiv:1603.04467}, 2016.

\bibitem{ioffe2015batch}
Sergey Ioffe and Christian Szegedy.
\newblock Batch normalization: Accelerating deep network training by reducing
  internal covariate shift.
\newblock In {\em International Conference on Machine Learning}, pages
  448--456, 2015.

\bibitem{zeiler2014visualizing}
Matthew~D Zeiler and Rob Fergus.
\newblock Visualizing and understanding convolutional networks.
\newblock In {\em European conference on computer vision}, pages 818--833.
  Springer, 2014.

\end{thebibliography}

\end{document}